\documentclass[12pt]{article}
\usepackage{latexsym}
\usepackage{amsmath,amsbsy,amssymb}
\usepackage{eepic}
\usepackage{epsfig}

\newcommand{\dd}{\mbox{\rm d}}

\newcommand{\gam}{\gamma}

\newcommand{\tl}{\tilde}

\newcommand{\be}{\begin{equation}}
\newcommand{\bear}{\begin{eqnarray}}
\newcommand{\ear}{\end{eqnarray}}
\newcommand{\ee}{\end{equation}}
\newcommand{\lbl}{\label}
\newcommand{\bi}{\bibitem}
\newcommand{\ci}{\cite}

\newcommand{\vs}{\vspace}
\newcommand{\hs}{\hspace}

\begin{document}

\

\baselineskip .7cm 

\vs{15mm}

\begin{center}

{\LARGE \bf  Pais-Uhlenbeck Oscillator with a Benign Friction Force}

\vs{3mm}

Matej Pav\v si\v c

Jo\v zef Stefan Institute, Jamova 39,
1000 Ljubljana, Slovenia

e-mail: matej.pavsic@ijs.si

\vs{6mm}

{\bf Abstract}
\end{center}

\baselineskip .5cm 

{\small
It is shown that the Pais-Uhlenbeck oscillator with damping, considered by
Nesterenko, is a special case of a more general oscillator that has not only
a first order, but also a third order friction term. If the
corresponding damping constants, $\alpha$ and $\beta$, are both positive and
below certain critical values, then
the system is stable. In particular, if $\alpha =-\beta$, then we have the
unstable Nesterenko's oscillator.

\section{Introduction}

\normalsize 

\baselineskip .59cm 

Pais-Uhlenbeck (PU) oscillator\,\ci{PaisUhlenbeck} is a toy model for higher
derivative theories.
The latter theories are very important for quantum gravity, but, because of the
presence of negative energies, they are generally considered as very problematic,
if not completely unsuitable for physics. Negative energies arise from the
wrong signs of certain terms in the Ostrogradsky Hamiltonian. In a quantized
theory, such wrong signs can manifest themselves in the presence of ghost
states\,\ci{Hawking} that break unitarity. With an alternative
quantization procedure, based
on a different choice of vacuum\,\ci{Jackiw,PavsicPseudoharm,Woodard}, one has negative
energy states, just as in the classical higher derivative theory, and no ghost
states.

Several authors have argued that the presence of negative energies in
PU oscillator does not lead to inconsistencies \,\ci{NegativeOK} (see also
Ref.\,\ci{NesterenkoRigid}). Those arguments
hold for a free oscillator, and are no longer valid if one includes
an interaction term that couples positive and negative energy degrees
of freedom.
The interacting PU oscillator has to be analyzed afresh.
In Refs.\,\ci{Carroll}--\ci{SmilgaStable} it has been found that for small
initial velocities and coupling constants there exist islands of stability.
Moreover, an example of an unconditionally stable interacting system was
found\,\ci{SmilgaStable}. This system, which is a non linear extension of the
PU oscillator, is a close relative of a
supersymmetric higher-derivative system\,\ci{Robert}.
Further, if to the ordinary, linear, PU oscillator we add a
self-interaction term that is bounded from below and from above, such as
$\frac{1}{4}\, {\rm sin}^4 \, x$,
then, as shown in Ref.\,\ci{PavsicPUstable}, such a system is
stable for any value of initial velocity, and is thus
an  example of a viable higher derivative theory.

But there remains an important issue that has to be resolved. Every physical
system in contact with an environment undergoes dissipative forces. An ordinary
oscillator is subjected to a damping force that exponentially diminishes
the amplitude of oscillations. For the PU oscillator, this could be different.
Indeed, according to Nesterenko\,\ci{Nesterenko}, the PU oscillator with
an external friction force undergoes an exponential instability:
the amplitude grows into infinity.

In this paper it will be shown that the friction force, considered by Nesterenko,
is a special case of a more general friction force, that in general does not
cause the exponential instability. Stability of such a system is also preserved
in the presence of an external time dependent force. 

\section{Pais-Uhlenbeck Oscillator with Damping}

Without damping, the Pais-Uhlenbeck oscillator satisfies the following forth order
equation of motion:
\be
    \left (\frac{\dd^2}{\dd t^2}+\omega_2^2 \right ) 
    \left (\frac{\dd^2}{\dd t^2}+\omega_1^2 \right ) x = 0.
\lbl{2.1} 
\ee
The latter equation can be generalized to include damping terms:
 \be
    \left (\frac{\dd^2}{\dd t^2}+ 2 \beta \frac{\dd}{\dd t}+\omega_2^2 \right ) 
    \left (\frac{\dd^2}{\dd t^2}+2 \alpha \frac{\dd}{\dd t}+\omega_1^2 \right ) x 
    = 0.
\lbl{2.2} 
\ee  
Explicitly we thus have
\be
    x^{(4)} + 2 (\alpha+\beta){\dddot x}
    +(\omega_1^2+\omega_2^2 +4 \alpha \beta) {\ddot x}
    +2 (\omega_1^2 \beta + \omega_2^2 \alpha) {\dot x} + \omega_1^2 \omega_2^2 x =0.
\lbl{2.3}
\ee

If $\alpha=-\beta$ we obtain the equation
\be
    x^{(4)} 
    +(\omega_1^2+\omega_2^2 -4 \beta^2) {\ddot x}
    +2 \beta (\omega_1^2 - \omega_2^2) {\dot x} + \omega_1^2 \omega_2^2\, x =0,
\lbl{2.4}
\ee    
which can be written in the form
\be
    x^{(4)} 
    +(\Omega_1^2+\Omega_2^2) {\ddot x}
    +2 \gamma {\dot x} + \Omega_1^2 \Omega_2^2 \, x =0,
\lbl{2.5}
\ee    
where $\gam = \beta (\omega_1^2 - \omega_2^2)$. Here
\be
    \Omega_1^2+\Omega_2^2 = \omega_1^2+\omega_2^2 - 4 \beta^2,
\lbl{2.6}
\ee
\be
     \Omega_1^2 \Omega_2^2 = \omega_1^2 \omega_2^2,
\lbl{2.7}
\ee
with the solution
\be
    \Omega_{1,2}^2 = \frac{1}{2} \left [ \omega_1^2 + \omega_2^2 - 4 \beta^2
      \pm \sqrt{(\omega_1^2 + \omega_2^2 - 4 \beta^2)^2- 4 \omega_1^2 \omega_2^2}
      \right ] .
\lbl{2.8}
\ee
Eq.\,(\ref{2.5}) is just the equation for the Pais-Uhlenbeck oscillator in
the presence of a friction force, considered
by Nesterenko\,\ci{Nesterenko}.

The general solution of Eq.\,(\ref{2.2}) is
\be
   x = {\rm e}^{- \alpha t} \left (C_1 {\rm e}^{t \sqrt{\alpha^2-\omega_1^2}}
       +C_2 {\rm e}^{- t \sqrt{\alpha^2-\omega_1^2}} \right )
     +{\rm e}^{- \beta t} \left (C_3 {\rm e}^{t \sqrt{\beta^2-\omega_2^2}}
       +C_4 {\rm e}^{- t \sqrt{\beta^2-\omega_2^2}} \right )
\lbl{2.9}
\ee
If $\alpha^2 < \omega_1^2$, $\beta^2 < \omega_2^2$, this is oscillatory
function, and if $\alpha$ and $\beta$ are both positive, the amplitude
of oscillations exponentially decreases.

In particular, if $\alpha =- \beta$, the solution of (\ref{2.2}) is
\be
x = {\rm e}^{\beta t}\left (
   C_1 {\rm e}^{t \sqrt{\beta^2-\omega_1^2}} 
   +C_2 {\rm e}^{-t \sqrt{\beta^2-\omega_1^2}} \right )
    +{\rm e}^{- \beta t} \left ( C_3 {\rm e}^{t \sqrt{\beta^2-\omega_2^2}}
       +C_4 {\rm e}^{- t \sqrt{\beta^2-\omega_2^2}} \right )
\lbl{2.10}
\ee
For $\beta^2 < \omega_1^2, \omega_2^2$, the $x(t)$ is oscillating function consisting
of a part with exponential growth, and a part with exponential damping.
Such behavior was found by Nesterenko, using a perturbative solution of
Eq.\,(\ref{2.5}). But as we see here, Eq.\,(\ref{2.5}) can be solved exactly
through the steps (\ref{2.2})--(\ref{2.8}), and by taking $\alpha=-\beta$.
Since Eq.\,(\ref{2.5}) is equivalent to the system of two oscillators
with the damping constants of opposite signs, it describes
an unstable system.

In general, for positive $\alpha \neq \beta$, Eq.\,(\ref{2.3}) has stable
solutions, provided that $|\alpha|$,
$|\beta|$ are sufficiently small, so that all terms are oscillating and
damped by ${\rm}^{-\alpha t}$ and ${\rm e}^{-\beta t}$.

\section{Presence of an arbitrary external force}

To the right hand side of the homogeneous equation (\ref{2.2}) we can
add an arbitrary time dependent force $f(t)$:    
 \be
    \left (\frac{\dd^2}{\dd t^2}+ 2 \beta \frac{\dd}{\dd t}+\omega_2^2 \right ) 
    \left (\frac{\dd^2}{\dd t^2}+2 \alpha \frac{\dd}{\dd t}+\omega_1^2 \right ) x 
    = f(t).
\lbl{3.1} 
\ee

In the absence of damping, $\alpha=\beta=0$, the general solution to the latter
equation can be expressed as\,\ci{Nesterenko}
\be
     x(t)= x_0 (t) +\int_{-\infty}^{\infty} G(t-t') f(t') \dd t' .
\lbl{3.2}
\ee
Here $x_0 (t)$ is the general solution of the homogeneous equation (\ref{2.1})
\be
   x_0 (t) = C_1 \,{\rm cos}\,\omega_1 t + C_2\, {\rm sin}\,\omega_1 t +
           C_3 \,{\rm cos}\,\omega_2 t + C_4\, {\rm sin}\,\omega_2 t ,  
\lbl{3.2a}
\ee
and
\be
      G(t) = \frac{1}{\sqrt{2 \pi}} \int_{-\infty}^{\infty} 
      {\rm e}^{i \omega t} {\tl G} (\omega) \dd \omega ,
\lbl{3.3}
\ee
\be
      f(t) = \frac{1}{\sqrt{2 \pi}} \int_{-\infty}^{\infty} 
      {\rm e}^{i \omega t} {\tl f} (\omega) \dd \omega ,
\lbl{3.4}
\ee
where\,\ci{Nesterenko}
\be
   {\tl G} (\omega) = \frac{1}{\sqrt{2 \pi}(\omega_1^2 - \omega_2^2)}
   \left (\frac{1}{\omega^2 - \omega_1^2}-\frac{1}{\omega^2 - \omega_2^2}
   \right ) .
\lbl{3.5}
\ee
By inserting (\ref{3.5}) into (\ref{3.3}), we obtain for $t>0$
\be
   G(t) = \frac{1}{2(\omega_1^2 - \omega_2^2)}
   \left ( \frac{{\rm sin}\,\omega_2 t}{\omega_2} -
          \frac{{\rm sin}\,\omega_1 t}{\omega_1} \right ) .
\lbl{3.6}
\ee
As an example let us first consider the force
\be
   f(t) = a \,{\rm cos}\,\omega_1 t + b\, {\rm cos}\,\omega_2 t ,
\lbl{3.7}
\ee
to which there corresponds the spectral density 
\be
{\tl f}(\omega)=\frac{a \sqrt{2 \pi}}{2}[\delta(\omega-\omega_1)
+\delta(\omega+\omega_1)] + \frac{b \sqrt{2 \pi}}{2}[
\delta(\omega - \omega_2)+\delta(\omega + \omega_2)].
\lbl{3.7a}
\ee
Then Eq.\,(\ref{3.2}) gives
   $$x(t) = x_0 (t) - \frac{1}{2 \omega_1 \omega_2 (\omega_1^2 - \omega_2^2)^2}
   \Bigl[ (\omega_1^2 - \omega_2^2) (a \omega_2 t \,{\rm sin}\,\omega_1 t
   - b \omega_1 t \,{\rm sin}\,\omega_2 t)  \hs{2cm}$$
\be   
  \hs{2cm} + 2 (a-b) \omega_1 \omega_2 ({\rm cos}\,\omega_1 t - {\rm cos}\,\omega_2 t)
   \Bigr]
\lbl{3.8}
\ee
The same solution of Eq.\,(\ref{3.1}) can be obtained also by using
the Mathematica command DSolve.

The amplitude in Eq.\,(\ref{3.8}) increases linearly with $t$. This was the
case without damping. If we include damping, we find the following general
solution of Eq.\,(\ref{3.1}):
\be
 x(t)=   x_0 (t) + \frac{-2 \beta \omega_1 a{\rm cos}\, \omega_1 t -
    a (\omega_1^2-\omega_2^2) {\rm sin}\,\omega_1 t}
    {2 \alpha \omega_1 [4 \beta^2 \omega_1^2 +(\omega_1^2-\omega_2^2)^2]}
    + \frac{-2\alpha \omega_2 b{\rm cos}\, \omega_2 t +
    b (\omega_1^2-\omega_2^2) {\rm sin}\,\omega_2 t}
    {2 \beta \omega_2 [4 \alpha^2 \omega_2^2 +(\omega_1^2-\omega_2^2)^2]} ,
\lbl{3.8a}
\ee
where $x_0 (t)$ is now the general solution of the homogeneous
equation (\ref{2.2}) (see Eq.\,(\ref{2.9})).
For positive $\alpha$ and $\beta$, satisfying $\alpha^2<\omega_1^2$,
$\beta^2<\omega_2^2$, the function $x(t)$ has the oscillating exponentially
decreasing part $x_0(t)$ (Eq.\,(\ref{2.9}), and the  oscillating part due to the
external force (\ref{3.7}). Notice that now, differently than in
Eq.\,(\ref{3.8}), the amplitude does not linearly
increase with $t$.

\setlength{\unitlength}{.8mm}

\begin{figure}[ht]
\hs{3mm} \begin{picture}(120,100)(10,-5)

\put(50,56){\epsfig{file=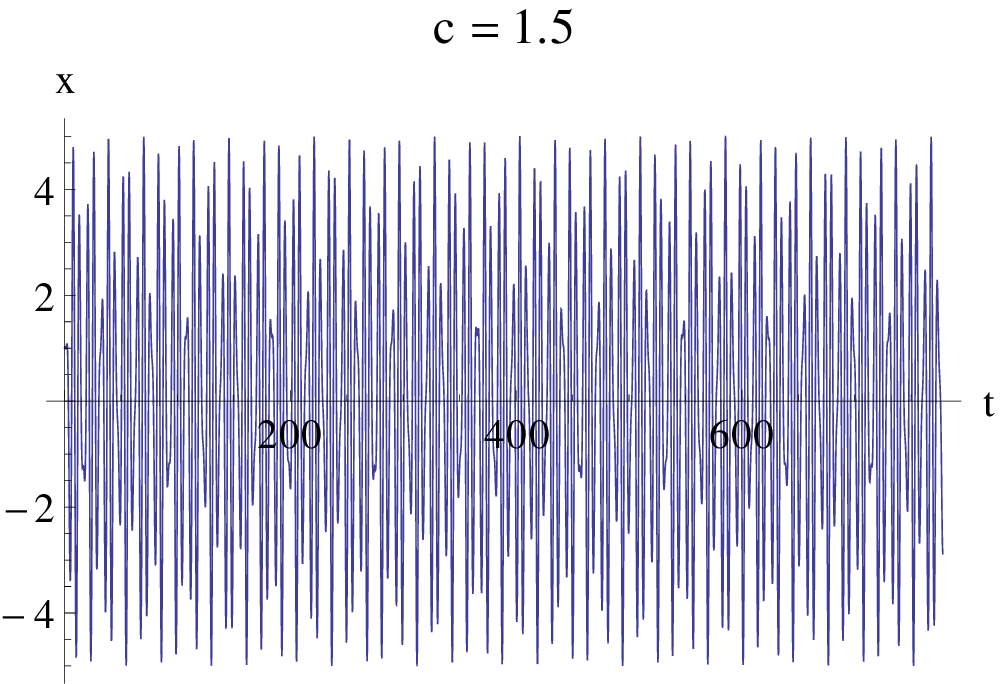,width=60mm}}
\put(47,0){\epsfig{file=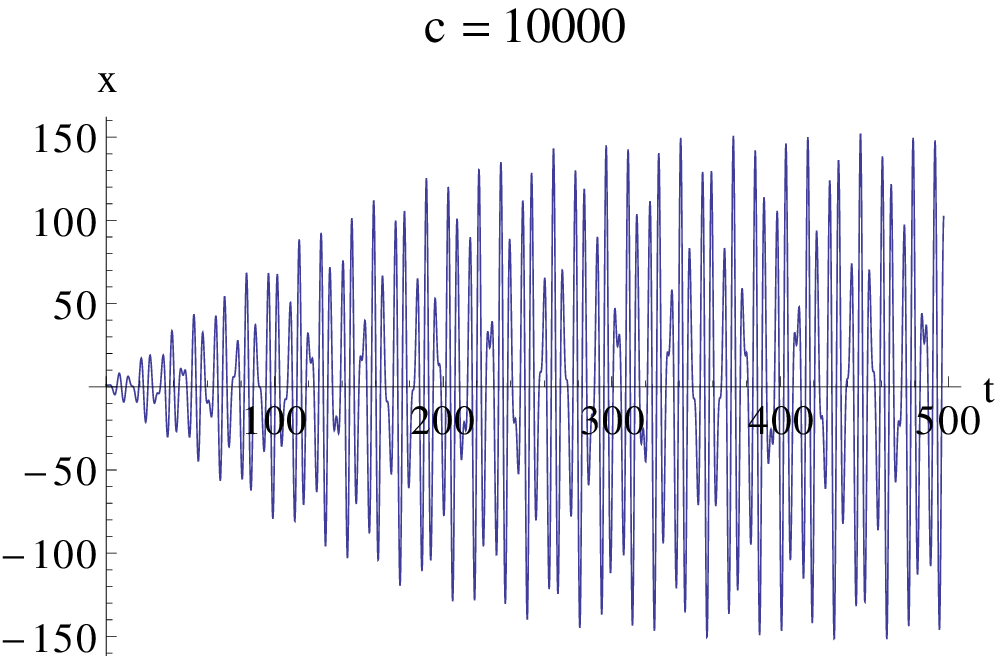,width=60mm}}

\end{picture}

\caption{\footnotesize Solution of the undamped Pais-Uhlenbeck oscillator
($\alpha=\beta=0$) in the presence
of an external force with the spectral density localized around $\omega_1^2=1$
and $\omega_2^2=1.5$ according to (\ref{3.9}) for two different values of the width
parameter $c$. We took the constants $a=b=1$, and the initial conditions
$x(0)=1$, ${\dot x}(0)=0.2$, ${\ddot x}(0)=-0.7$, ${\dddot x}(0)=0.5$.}

\end{figure}

More generally, if the spectral density is localized around $\omega_1^2$
and $\omega_2^2$ according to
\be
    {\tl f} (\omega) = \frac{a \sqrt{2 \pi}}{2} \sqrt{\frac{c}{\pi}} 
    \left ( {\rm e}^{-c(\omega - \omega_1)^2}
    +{\rm e}^{-c(\omega + \omega_1)^2} \right )
  + \frac{b \sqrt{2 \pi}}{2} \sqrt{\frac{c}{\pi}} 
  \left ( {\rm e}^{-c(\omega - \omega_2)^2} 
  + {\rm e}^{-c(\omega + \omega_2)^2} \right ),
\lbl{3.9}
\ee
then
\be
     f(t) = {\rm e}^{-\frac{t^2}{4 c}}
     (a \,{\rm cos}\,\omega_1 t + b\, {\rm cos}\,\omega_2 t ) ,
\lbl{3.10}
\ee
and solutions to Eq.\,(\ref{3.1}) are stable, oscillating functions
even in the absence of damping. This can be verified by solving Eq.\,(\ref{3.1})
numerically, using the command NDSolve in Mathematica. Examples of solutions
are given in Fig.\,1.
We see that nothing unphysical happens with the classical displacement $x(t)$.
It displays decent oscillatory behavior. Quantum behavior of the propagator
has been recently investigated by Ilhan and Kovner\,\ci{Ilhan}. 

\section{Conclusion}

We have clarified the important point raised by Nesterenko and found that
the Pais-Uhlenbeck oscillator with external friction force is not necessarily
unstable. It can be stable, because in general, the Pais-Uhlenbeck oscillator
has not only one, but two damping terms: a term with the first and
a term with the third derivative
of the displacement $x(t)$ with respect to the time $t$. If the corresponding
damping constants, $\alpha$ and $\beta$, are both positive and lower than
the critical values determined by $\omega_1^2$, $\omega_2^2$, then the system
is stable in the sense that it oscillates with exponentially decreasing
amplitude. In particular, if $\alpha = -\beta$, then the third order
term vanishes and we have the
oscillator that has only the first derivative damping term, considered by
Nesterenko\,\ci{Nesterenko}. Such "damping" term causes the exponential growth
of the oscillator's amplitude. Nesterenko's conclusion that the theories
with higher derivatives suffer exponential instability holds only in the latter
particular case. In general, such theories can be stable, because the third
order damping term provides the mechanism that prevents the exponential
instability.

We have also analysed the Pais-Uhlenbeck oscillator that experiences an
arbitrary external force, $f(t)$. For the force whose spectral density
is sharply localized around $\omega_1^2$ and $\omega_2^2$,
we have found the exact general solution whose
amplitude linearly increases with time if $\alpha=\beta=0$, and decently
oscillates if $\alpha>0$, $\beta>0$. For a force whose spectral density
has Gaussian (and not the physically unrealistic $\delta$-like)
distribution around $\omega_1^2$ and $\omega_2^2$ , we obtain
stable, oscillating solutions even in the absence of damping.

\vs{2mm}

\centerline{\bf Acknowledgment}

This work has been supported by the Slovenian Research Agency.

\end{document}